\documentclass[11pt,]{article}
\usepackage{lmodern}
\usepackage{amssymb,amsmath}
\usepackage{ifxetex,ifluatex}
\usepackage{fixltx2e} % provides \textsubscript
\ifnum 0\ifxetex 1\fi\ifluatex 1\fi=0 % if pdftex
  \usepackage[T1]{fontenc}
  \usepackage[utf8]{inputenc}
\else % if luatex or xelatex
  \ifxetex
    \usepackage{mathspec}
  \else
    \usepackage{fontspec}
  \fi
  \defaultfontfeatures{Ligatures=TeX,Scale=MatchLowercase}
\fi
\IfFileExists{upquote.sty}{\usepackage{upquote}}{}
\IfFileExists{microtype.sty}{%
\usepackage{microtype}
\UseMicrotypeSet[protrusion]{basicmath} % disable protrusion for tt fonts
}{}
\usepackage[margin=1in]{geometry}
\usepackage{hyperref}
\hypersetup{unicode=true,
            pdftitle={merlin - a unified modelling framework for data analysis and methods development in Stata},
            pdfauthor={University of Leicester},
            pdfborder={0 0 0},
            breaklinks=true}
\usepackage{natbib}
\usepackage{graphicx,grffile}
\makeatletter
\def\maxwidth{\ifdim\Gin@nat@width>\linewidth\linewidth\else\Gin@nat@width\fi}
\def\maxheight{\ifdim\Gin@nat@height>\textheight\textheight\else\Gin@nat@height\fi}
\makeatother
\setkeys{Gin}{width=\maxwidth,height=\maxheight,keepaspectratio}
\IfFileExists{parskip.sty}{%
\usepackage{parskip}
}{% else
\setlength{\parindent}{0pt}
\setlength{\parskip}{6pt plus 2pt minus 1pt}
}
\providecommand{\tightlist}{%
  \setlength{\itemsep}{0pt}\setlength{\parskip}{0pt}}
\ifx\paragraph\undefined\else
\let\oldparagraph\paragraph
\renewcommand{\paragraph}[1]{\oldparagraph{#1}\mbox{}}
\fi
\ifx\subparagraph\undefined\else
\let\oldsubparagraph\subparagraph
\renewcommand{\subparagraph}[1]{\oldsubparagraph{#1}\mbox{}}
\fi

\let\rmarkdownfootnote\footnote%
\def\footnote{\protect\rmarkdownfootnote}

\usepackage{titling}

\newcommand{\subtitle}[1]{
  \posttitle{
    \begin{center}\large#1\end{center}
    }
}

  \title{\texttt{merlin} - a unified modelling framework for data analysis and
methods development in \texttt{Stata}}
  \pretitle{\vspace{\droptitle}\centering\huge}
  \posttitle{\par}
\subtitle{Michael J. Crowther}
  \author{University of Leicester}
  \preauthor{\centering\large\emph}
  \postauthor{\par}
  \predate{\centering\large\emph}
  \postdate{\par}
  \date{\href{mailto:michael.crowther@le.ac.uk}{\nolinkurl{michael.crowther@le.ac.uk}}}

\begin{document}
\maketitle
\begin{abstract}
\texttt{merlin} can do a lot of things. From simple stuff, like fitting
a linear regression or a Weibull survival model, to a three-level
logistic mixed effects model, or a multivariate joint model of multiple
longitudinal outcomes (of different types) \emph{and} a recurrent event
\emph{and} survival with non-linear effects\ldots{}the list is rather
endless. \texttt{merlin} can do things I haven't even thought of yet.
I'll take a single dataset, and attempt to show you the full range of
capabilities of \texttt{merlin}, and discuss some future directions for
the implementation in \texttt{Stata}.
\end{abstract}

\begin{center}
Draft v1: \today
\end{center}

\section{Introduction}\label{introduction}

\texttt{gsem}'s introduction in Stata 14 brought an extremely broad
class of mixed effects models (among other things), and most importantly
(from my perspective), is that \texttt{gsem} is fast. \texttt{gsem} has
full analytic derivatives for any model that you fit, including a model
with any number of levels and random effects at each level. Of course,
\texttt{gllamm} came along before \texttt{gsem}, providing a very
flexible framework in Stata for modelling data
\citep{RHgllamm, Rabe-Hesketh2005}. But inevitably, every program is
limited to certain families of distributions, and complexity of the
linear predictors, and of course computational speed.

Much of my previous work has centred on joint longitudinal-survival
analysis, implemented in the \texttt{stjm} command
\citep{Crowther2012sj}, which can fit a joint model for a continuous,
repeatedly measured biomarker, and a single time-to-event. There is
flexibility, in that you can use splines or polynomials to model the
biomarker over time, and there's lots of choices for your survival
model, including spline based approaches. The main benefit of
\texttt{stjm} is the added flexibility in how to link the two submodels,
through things like the expected value, or derivatives of it, which are
commonly used in the joint model literature. Given this starting point,
the natural extension is to allow for more than one biomarker, recurrent
events, competing risks\ldots{}the list goes on. I've also released
\texttt{stmixed} in Stata \citep{Crowther2014a}, for two-level
parametric survival models, in particular for the Royston-Parmar spline
based model, and \texttt{stgenreg} for general hazard based regression
\citep{stgenreg, Crowther2014}, where the user could specify their own
functional form for the hazard function, and \texttt{stgenreg} would use
numerical quadrature to maximise the likelihood.

The aim of \texttt{merlin} was to bring together my previous programs,
but also provide a whole lot more. In this paper, I'll introduce the
core features and syntax of \texttt{merlin} and illustrate a variety of
models that can be fitted with it. More details on the methodological
framework can be found in \citet{Crowther2017c}. The paper is structured
as follows: In Section \ref{sec:merlin} I describe the top level syntax
of the \texttt{merlin} command and describe many of the options, and in
Section \ref{sec:eg} I detail some examples, showing the wide variety of
available models. I conclude the paper in Section \ref{sec:disc} with a
discussion of \texttt{merlin} and its potential.

\section{\texorpdfstring{The architecture of
\texttt{merlin}}{The architecture of merlin}}\label{the-architecture-of-merlin}

\label{sec:merlin}

\texttt{merlin} is designed to be as flexible and general as possible.
There is no real limit to what it can do, given that the user can
readily extend it. You can specify any number of outcome models, which
can be linked in any number of ways. Each outcome model,
\(i=1,\dots,M\), has a main complex predictor (you can have more, but
I'll leave that to another time), which is made up of additive
components, \(c=1,\dots,C\), where each component is made up of
multiplicative elements, \(e=1,\dots,E\).

\[
g_{i}(\mu_{i}(y_{i} | x,b)) = \sum_{c=1}^{C} \prod_{e=1}^{E} \psi_{ice}(x,b)
\]

where for the \(i^{th}\) model, \(y_{i}\) is the observed response,
\(x\) is the full design matrix, and \(b\) are the stacked multivariate
normal or \(t\)-distributed random effects. We have link function,
\(g_{i}()\), and expected value \(\mu_{i}\). \(\psi()\) defines
essentially an arbitrary function, of which some special cases are
defined in the next section.

\texttt{merlin} is estimated using maximum likelihood, with any random
effects integrated out using either adaptive or non-adaptive Gaussian
quadrature, or Monte Carlo integration. More details on the formulation
of the likelihood can be found in \citet{Crowther2017c}.

\subsection{\texorpdfstring{\texttt{merlin}
syntax}{merlin syntax}}\label{merlin-syntax}

\texttt{merlin\ (}\emph{\texttt{model1}}\texttt{)\ {[}(}\emph{\texttt{model2}}\texttt{){]}\ {[}...{]}\ {[},}\emph{\texttt{options}}\texttt{{]}}

\hspace{5ex}where the syntax of a \emph{\texttt{modeli}} is

\hspace{5ex}\texttt{{[}}\emph{\texttt{depvar}}\texttt{{]}\ {[}}\emph{\texttt{component1}}\texttt{{]}\ {[}}\emph{\texttt{component2}}\texttt{{]}\ {[}...{]}\ {[},}\emph{\texttt{model\_options}}\texttt{{]}}

\hspace{10ex}where the syntax of a \emph{\texttt{componentj}} is

\hspace{10ex}\emph{\texttt{element1}}\texttt{{[}\#}\emph{\texttt{element2}}\texttt{{]}{[}\#}\emph{\texttt{element3}}\texttt{{]}{[}...{]}{[}@}\emph{\texttt{real}}\texttt{{]}}

and each \emph{\texttt{elementk}} can take one of the forms described in
the next section. At the end of each component, you may optionally
specify a constraint on the parameter(s) of the associated component,
through use of the \texttt{@}.

\emph{\texttt{model\_options}} include:

\begin{itemize}
\tightlist
\item
  \texttt{family(}\emph{\texttt{fam}}\texttt{,}\emph{\texttt{fam\_options}}\texttt{)},
  where \emph{\texttt{fam}} can be one of:

  \begin{itemize}
  \tightlist
  \item
    \texttt{gaussian} - Gaussian distribution
  \item
    \texttt{bernoulli} - Bernoulli distribution
  \item
    \texttt{beta} - beta distribution
  \item
    \texttt{poisson} - Poisson distribution
  \item
    \texttt{ologit} - ordinal with logistic link
  \item
    \texttt{oprobit} - ordinal with probit link
  \item
    \texttt{exponential} - exponential distribution
  \item
    \texttt{weibull} - Weibull distribution
  \item
    \texttt{gompertz} - Gompertz distribution
  \item
    \texttt{gamma} - gamma distribution
  \item
    \texttt{rp} - Royston-Parmar model (restricted cubic spline on log
    cumulative hazard scale)
  \item
    \texttt{rcs} - restricted cubic splines on log hazard scale
  \item
    \texttt{user} - specify your own distribution \newline and
    \emph{\texttt{fam\_options}} include:
  \item
    \texttt{failure(varname)} - event indicator for a survival model
  \item
    \texttt{ltruncated(varname)} - left-truncation/delayed entry times
    for survival model
  \item
    \texttt{llfunction(}\emph{\texttt{func\_name}}\texttt{)} - name of a
    \texttt{Mata} function which returns your user-defined
    log-likelihood contribution
  \item
    \texttt{hazard(}\emph{\texttt{func\_name}}\texttt{)} - name of a
    \texttt{Mata} function which returns your user-defined hazard
    function
  \item
    \texttt{chazard(}\emph{\texttt{func\_name}}\texttt{)} - name of a
    \texttt{Mata} function which returns your user-defined cumulative
    hazard function
  \item
    \texttt{nap(\#)} - estimate \texttt{\#} ancillary parameters, which
    may be called in user-defined functions
  \end{itemize}
\item
  \texttt{timevar(varname)} - specifies the variable which contains
  time; this is required to specify time-dependent effects. Generally
  within a survival model, time must be explicitly handled by
  \texttt{merlin}. \texttt{timevar()} will be matched against any
  elements (see the next section) which may use it, to make sure it is
  handled correctly.
\end{itemize}

There are many other suboptions available in \texttt{merlin}, which are
fully documented in the \texttt{help} files.

\subsubsection{Element types}\label{element-types}

One of the fundamental flexibilities of \texttt{merlin} is the variety
of elements that can be used within your model.

\begin{itemize}
\tightlist
\item
  \texttt{varname} - an independent variable in your dataset
\item
  \texttt{rcs(varname\ ,}\emph{\texttt{opts}}\texttt{)} - a restricted
  cubic spline function

  \begin{itemize}
  \tightlist
  \item
    \texttt{df(\#)} - degrees of freedom (\# of internal knots + 1).
    Knots are placed at evenly spaced centiles of \texttt{varname}, with
    boundary knots assumed to be the minimum and maximum of
    \texttt{varname}.
  \item
    \texttt{knots(numlist)} - specifies the knots, including the
    boundary knots.
  \item
    \texttt{log} - create splines of \texttt{log(varname)} instead of
    the default untransformed \texttt{varname}
  \item
    \texttt{orthog} - apply Gram-Schmidt orthogonalisation; can improve
    convergence
  \item
    \texttt{event} - used in conjunction with \texttt{df()}, specifies
    that internal knot locations are based on centiles of only
    observations of \texttt{varname} that experienced the survival event
    specified in \texttt{failure()}
  \item
    \texttt{offset(}\emph{\texttt{varname}}\texttt{)} - specifies an
    offset, to be added to \texttt{varname} before the \texttt{fp()}
    function is built
  \end{itemize}
\item
  \texttt{bs(varname\ ,}\emph{\texttt{opts}}\texttt{)} - a B-spline
  function

  \begin{itemize}
  \tightlist
  \item
    \texttt{df(\#)} - specifies the degrees of freedom (not strictly
    speaking) for the spline function, which allows you to specify
    internal knots at equally spaced centiles, instead of using
    \texttt{knots()}. \texttt{df()} is consistent with \texttt{rcs()}
    elements in how internal knots are chosen.
  \item
    \texttt{knots(numlist)} - specifies the internal knots, must be in
    ascending order.
  \item
    \texttt{bknots(numlist)} - specifies the lower and upper boundary
    knot locations. Must be in ascending order. Default is the minimum
    and maximum of varname.
  \item
    \texttt{intercept} - includes the intercept basis function, which by
    default is not included.
  \item
    \texttt{log} - create splines of \texttt{log(varname)} instead of
    the default untransformed \texttt{varname}
  \item
    \texttt{event} - used in conjunction with \texttt{df()}, specifies
    that internal knot locations are based on centiles of only
    observations of \texttt{varname} that experienced the survival event
    specified in \texttt{failure()}
  \item
    \texttt{offset(}\emph{\texttt{varname}}\texttt{)} - specifies an
    offset, to be added to \texttt{varname} before the \texttt{fp()}
    function is built
  \end{itemize}
\item
  \texttt{fp(varname\ ,}\emph{\texttt{opts}}\texttt{)} - a fractional
  polynomial function (of order 1 or 2), where opts include:

  \begin{itemize}
  \tightlist
  \item
    \texttt{powers(}\emph{\texttt{numlist}}\texttt{)} - powers up to a
    second-degree FP function, each of which can be one of
    \texttt{(-2,-1,-0.5,0,0.5,1,2,3)}
  \item
    \texttt{offset(}\emph{\texttt{varname}}\texttt{)} - specifies an
    offset, to be added to \texttt{varname} before the \texttt{fp()}
    function is built
  \end{itemize}
\item
  \texttt{mf(}\emph{\texttt{func\_name}}\texttt{)} - a user-defined
  \texttt{Mata} function (see the section on utility functions)
\item
  \texttt{M\#{[}cluster\ level{]}} - a random effect, defined at the
  \texttt{cluster\ level}, for example, \texttt{M1{[}centre{]}} defines
  a random intercept at the \texttt{centre} level, and
  \texttt{M2{[}centre\textgreater{}id{]}} defines a random intercept at
  the \texttt{id} level.
\item
  \texttt{EV{[}depvar/\#{]}} - the expected value of an outcome model
\item
  \texttt{dEV{[}depvar/\#{]}} - the first derivative with respect to
  time of the expected value of an outcome model
\item
  \texttt{d2EV{[}depvar/\#{]}} - the second derivative with respect to
  time of the expected value of an outcome model
\item
  \texttt{iEV{[}depvar/\#{]}} - the integral with respect to time of the
  expected value of an outcome model
\item
  \texttt{XB{[}depvar/\#{]}} - the expected value of a complex predictor
\item
  \texttt{dXB{[}depvar/\#{]}} - the first derivative with respect to
  time of the expected value of a complex predictor
\item
  \texttt{d2XB{[}depvar/\#{]}} - the second derivative with respect to
  time of the expected value of a complex predictor
\item
  \texttt{iXB{[}depvar/\#{]}} - the integral with respect to time of the
  expected value of a complex predictor
\end{itemize}

\subsubsection{Utility functions}\label{utility-functions}

On one hand, \texttt{merlin} is an engine to estimate many standard
models, such as multivariate hierarchical models. On the other, it's a
platform on which to extend and build on. One of the main design
features of \texttt{merlin} is the ease at which it can be extended and
added to by the user. This makes it extremely useful for methodological
research. Utility functions can be used in two main settings.

\begin{enumerate}
\def\labelenumi{\arabic{enumi}.}
\tightlist
\item
  \texttt{family(user)} - when passing your own log-likelihood function,
  or hazard and/or cumulative hazard function, utility functions can be
  used within the Mata function to provide access to anything you may
  need from the \texttt{merlin} model object
\item
  \texttt{mf()} - when specifying your own element type, you can also
  call any of the utility functions
\end{enumerate}

Available utility functions include:

\begin{itemize}
\tightlist
\item
  \texttt{merlin\_util\_depvar(M)} - returns the dependent variable for
  the current model. This will be a matrix with two columns if
  \texttt{failure()} is specified, and three columns if
  \texttt{ltruncated()} has been specified.
\item
  \texttt{merlin\_util\_xzb(M\ ,\ \textbar{}\ real\ colvector\ t)} -
  returns the complex predictor for the current model, optionally
  evaluated at time points \texttt{t}
\item
  \texttt{merlin\_util\_xzb\_deriv(M\ ,\ \textbar{}\ real\ colvector\ t)}
  - returns d/dt of the complex predictor for the current model,
  optionally evaluated at time points \texttt{t}
\item
  \texttt{merlin\_util\_xzb\_deriv2(M\ ,\ \textbar{}\ real\ colvector\ t)}
  - returns d2/dt2 of the complex predictor for the current model,
  optionally evaluated at time points \texttt{t}
\item
  \texttt{merlin\_util\_xzb\_integ(M\ ,\ \textbar{}\ real\ colvector\ t)}
  - returns the integral with respect to time of the complex predictor
  for the current model, optionally evaluated at time points \texttt{t}
\item
  \texttt{merlin\_util\_expval(M\ ,\ \textbar{}\ real\ colvector\ t)} -
  returns the expected value of the response for the current model,
  optionally evaluated at time points \texttt{t}
\item
  \texttt{merlin\_util\_expval\_deriv(M\ ,\ \textbar{}\ real\ colvector\ t)}
  - returns d/dt of the expected value of the response for the current
  model, optionally evaluated at time points \texttt{t}
\item
  \texttt{merlin\_util\_expval\_deriv2(M\ ,\ \textbar{}\ real\ colvector\ t)}
  - returns d2/dt2 of the expected value of the response for the current
  model, optionally evaluated at time points \texttt{t}
\item
  \texttt{merlin\_util\_expval\_integ(M\ ,\ \textbar{}\ real\ colvector\ t)}
  - returns the integral of the expected value of the response for the
  current model, optionally evaluated at time points \texttt{t}
\item
  \texttt{merlin\_util\_ap(M,\#)}- returns the \texttt{\#}th ancillary
  parameter of the current model
\item
  \texttt{merlin\_util\_timevar(M)} - returns the time variable for the
  current model, which would've been specified using the
  \texttt{timevar(varname)} option
\end{itemize}

All of the utility function take as their first argument a
\texttt{transmorphic} object, in this case I have called it \texttt{M},
but you may call it anything you like. This contains the \texttt{merlin}
object, which should not be edited. All of the \texttt{xzb} or
\texttt{expval} utility functions have an equivalent \texttt{*\_mod()}
function, which allows you to specify an additional argument,
representing the model which you want to call, e.g.
\texttt{merlin\_util\_xzb\_mod(M,2)} will return the complex predictor
for the second model in your \texttt{merlin} statement. This can be
included in your function for the first, letting you link submodels.

\subsubsection{\texorpdfstring{\texttt{merlin}
postestimation}{merlin postestimation}}\label{merlin-postestimation}

\texttt{merlin} comes with a set of post-estimation tools available
through the \texttt{predict} function, with the standard syntax of:

\texttt{predict}
\emph{\texttt{newvarname}}\texttt{,}\hspace{1ex}\emph{\texttt{statistic}}
\texttt{{[}}\emph{\texttt{options}}\texttt{{]}}

where \emph{\texttt{statistic}} includes:

\begin{itemize}
\tightlist
\item
  \texttt{mu} - expected value of the response
\item
  \texttt{eta} - expected value of the complex predictor
\item
  \texttt{hazard} - hazard function
\item
  \texttt{chazard} - cumulative hazard function
\item
  \texttt{survival} - survival function
\item
  \texttt{cif} - cumulative incidence function
\item
  \texttt{rmst} - restricted mean survival time (integral of
  \texttt{survival})
\item
  \texttt{timelost} - time lost due to the event (integral of
  \texttt{cif})
\end{itemize}

and \emph{\texttt{options}} include:

\begin{itemize}
\tightlist
\item
  \texttt{outcome(\#)} - specifies the model to predict for; default is
  \texttt{outcome(1)}
\item
  \texttt{fixedonly} - calculate prediction based only on the fixed
  effects
\item
  \texttt{marginal} - calculate prediction integrating out all random
  effects, i.e.~the population-averaged prediction
\item
  \texttt{at(}\emph{\texttt{at\_spec}}\texttt{)} - specify covariate
  patterns at which to calculate the \emph{\texttt{statistic}} at, e.g.
  \texttt{at(trt\ 1\ age\ 54)}
\item
  \texttt{timevar(}\emph{\texttt{varname}}\texttt{)} - specify a
  variable which contains timepoints at which to calculate the
  \emph{\texttt{statistic}} at
\item
  \texttt{ci} - calculate confidence intervals, using the delta method
  through \texttt{predictnl}
\item
  \texttt{causes(}\emph{\texttt{numlist}}\texttt{)} - specifies which
  models contribute to the calculation; for use in competing risks
  models
\end{itemize}

\section{Examples}\label{examples}

\label{sec:eg}

Given how varied and arguably complex the syntax can appear, the easiest
way to get to grips with \texttt{merlin} is through some examples.
Throughout this section I will use a single dataset
\citep{Murtaugh1994}. It's a commonly used dataset from the joint
longitudinal-survival literature, and will serve to illustrate many
different analysis techniques, culminating in a detailed, complex
multivariate hierarchical model. First I load the data, which you can
get from my website,

\begin{verbatim}
use "https://www.mjcrowther.co.uk/data/jm_example.dta", clear
\end{verbatim}

The dataset consists of information on 312 patients with primary biliary
cirrhosis, of which 140 died during a maximum follow-up of 14.3 years.
Covariates of interest include serum bilirubin and prothrombin index,
both markers of liver performance, and treatment allocation,
\texttt{trt}. Patients were randomised to either D-penicillamine or a
placebo. In all analyses, I will work with the log of serum bilirubin,
stored in \texttt{logb}. The data structure looks as follows:

\begin{verbatim}
. list id stime died trt logb pro time if inlist(id,1,2), sepby(id) noobs

  +------------------------------------------------------------------+
  | id     stime   died         trt        logb   prothr~n      time |
  |------------------------------------------------------------------|
  |  1   1.09517      1   D-penicil    2.674149       12.2         0 |
  |  1         .      .   D-penicil    3.058707       11.2   .525682 |
  |------------------------------------------------------------------|
  |  2   14.1523      0   D-penicil    .0953102       10.6         0 |
  |  2         .      .   D-penicil   -.2231435         11   .498302 |
  |  2         .      .   D-penicil           0       11.6   .999343 |
  |  2         .      .   D-penicil    .6418539       10.6   2.10273 |
  |  2         .      .   D-penicil    .9555114       11.3   4.90089 |
  |  2         .      .   D-penicil    1.280934       11.5   5.88928 |
  |  2         .      .   D-penicil    1.435084       11.5   6.88588 |
  |  2         .      .   D-penicil    1.280934       11.5    7.8907 |
  |  2         .      .   D-penicil    1.526056       11.5   8.83255 |
  +------------------------------------------------------------------+
\end{verbatim}

\texttt{merlin}, just like \texttt{gsem}, treats models in wide format,
but observations within a model in long format. Hence, our survival
outcome variables, \texttt{stime} and \texttt{died}, must only have one
observation per \texttt{id}. If a patient had more than one row of data
for their survival outcome, then we could fit a recurrent event
model\ldots{}but that's a tangent for another time. Speaking of time,
our \texttt{time} variable records the times at which the biomarkers
\texttt{logb} and \texttt{prothrombin} were recorded.

\subsection{Linear mixed effects
regression}\label{linear-mixed-effects-regression}

I will start with a very simple model, a linear regression of
\texttt{logb} against \texttt{time}, assuming a Gaussian response:

\begin{verbatim}
. merlin (logb time, family(gaussian))

Fitting full model:

Iteration 0:   log likelihood = -3339.4091  
Iteration 1:   log likelihood = -3044.5062  
Iteration 2:   log likelihood = -2962.0949  
Iteration 3:   log likelihood = -2961.4148  
Iteration 4:   log likelihood = -2961.4144  
Iteration 5:   log likelihood = -2961.4144  

Mixed effects regression model                  Number of obs     =      1,945
Log likelihood = -2961.4144
------------------------------------------------------------------------------
             |      Coef.   Std. Err.      z    P>|z|     [95% Conf. Interval]
-------------+----------------------------------------------------------------
logb:        |            
        time |   .0139443   .0081287     1.72   0.086    -.0019876    .0298763
       _cons |   .5594103   .0358095    15.62   0.000      .489225    .6295956
  sd(resid.) |   1.109201   .0177842                      1.074886    1.144611
------------------------------------------------------------------------------
\end{verbatim}

Let's add some flexibility by using the \texttt{rcs()} element, which
lets us model the change over time flexibly using restricted cubic
splines with three degrees of freedom, i.e.~three spline terms.

\begin{verbatim}
. merlin (logb rcs(time, df(3)), family(gaussian)), nolog
variables created for model 1, component 1: _cmp_1_1_1 to _cmp_1_1_3

Fitting full model:

Mixed effects regression model                  Number of obs     =      1,945
Log likelihood = -2960.7317
------------------------------------------------------------------------------
             |      Coef.   Std. Err.      z    P>|z|     [95% Conf. Interval]
-------------+----------------------------------------------------------------
logb:        |            
     rcs():1 |   .0636201    .065479     0.97   0.331    -.0647164    .1919567
     rcs():2 |   .0057442   .0115642     0.50   0.619    -.0169212    .0284096
     rcs():3 |  -.0012839   .0034269    -0.37   0.708    -.0080005    .0054327
       _cons |    .518356   .0540793     9.59   0.000     .4123626    .6243494
  sd(resid.) |   1.108811    .017778                      1.074509    1.144209
------------------------------------------------------------------------------
\end{verbatim}

If you prefer fractional polynomials or B-splines, then use the
\texttt{fp()} or \texttt{bs()} element types. Given our observations are
clustered within individuals, let's add a random intercept:

\begin{verbatim}
. merlin (logb rcs(time, df(3)) M1[id]@1, family(gaussian)), nolog
variables created for model 1, component 1: _cmp_1_1_1 to _cmp_1_1_3

Fitting fixed effects model:

Fitting full model:

Mixed effects regression model                  Number of obs     =      1,945
Log likelihood = -1871.1924
------------------------------------------------------------------------------
             |      Coef.   Std. Err.      z    P>|z|     [95% Conf. Interval]
-------------+----------------------------------------------------------------
logb:        |            
     rcs():1 |   .1157301   .0300614     3.85   0.000     .0568108    .1746494
     rcs():2 |  -.0047585   .0052904    -0.90   0.368    -.0151275    .0056104
     rcs():3 |   .0024164   .0015696     1.54   0.124    -.0006599    .0054928
      M1[id] |          1          .        .       .            .           .
       _cons |   .5311768   .0666663     7.97   0.000     .4005132    .6618403
  sd(resid.) |   .4866347   .0085298                      .4702005    .5036433
-------------+----------------------------------------------------------------
id:          |            
      sd(M1) |   1.099119   .0466923                       1.01131    1.194552
------------------------------------------------------------------------------
\end{verbatim}

I have added \texttt{M1{[}id{]}@1} to my complex predictor. This defines
a single normally distributed random effect, called \texttt{M1}, defined
at the \texttt{id} level. By default, any component within the complex
predictor will have an estimated coefficient. Given our model will
already estimate a fixed intercept, we want to constrain the random
effects coefficient to be 1, by specifying \texttt{@1} at the end. Let's
add a random linear slope, and also orthogonalise my spline terms:

\begin{verbatim}
. merlin (logb rcs(time, df(3) orthog) time#M2[id]@1 M1[id]@1, family(gaussian)), nolog
variables created for model 1, component 1: _cmp_1_1_1 to _cmp_1_1_3

Fitting fixed effects model:

Fitting full model:

Mixed effects regression model                  Number of obs     =      1,945
Log likelihood = -1531.5158
------------------------------------------------------------------------------
             |      Coef.   Std. Err.      z    P>|z|     [95% Conf. Interval]
-------------+----------------------------------------------------------------
logb:        |            
     rcs():1 |   .5469075   .0439678    12.44   0.000     .4607322    .6330827
     rcs():2 |  -.0364925   .0115305    -3.16   0.002    -.0590918   -.0138932
     rcs():3 |   .0115627   .0091666     1.26   0.207    -.0064034    .0295289
 time#M2[id] |          1          .        .       .            .           .
      M1[id] |          1          .        .       .            .           .
       _cons |   1.034689   .0702843    14.72   0.000     .8969347    1.172444
  sd(resid.) |   .3445818   .0066951                      .3317063     .357957
-------------+----------------------------------------------------------------
id:          |            
      sd(M1) |   1.011593   .0428349                      .9310276     1.09913
      sd(M2) |   .1814674   .0127782                      .1580739    .2083228
------------------------------------------------------------------------------
\end{verbatim}

Note by orthogonalising the splines, it changes the interpretation of
the intercept. I now have a component with more than one element,
\texttt{time\#M2{[}id{]}@1} - a variable \texttt{time} has been
interacted with a random effect called \texttt{M2} defined at the
\texttt{id} level, with its coefficient constrained to be 1. If I'd
called it \texttt{M1} again, then there would only be a single random
effect in my model, but used twice, which would not be sensible in this
case. By default, each level's random effects are assumed to have
\texttt{covariance(diagonal)}. We can relax this by estimating a
correlation:

\begin{verbatim}
. merlin (logb rcs(time, df(3) orthog) time#M2[id]@1 M1[id]@1, family(gaussian)),  ///
>                                       covariance(unstructured) nolog
variables created for model 1, component 1: _cmp_1_1_1 to _cmp_1_1_3

Fitting fixed effects model:

Fitting full model:

Mixed effects regression model                  Number of obs     =      1,945
Log likelihood =  -1518.484
------------------------------------------------------------------------------
             |      Coef.   Std. Err.      z    P>|z|     [95% Conf. Interval]
-------------+----------------------------------------------------------------
logb:        |            
     rcs():1 |   .5975687    .044098    13.55   0.000     .5111382    .6839992
     rcs():2 |  -.0377092    .011405    -3.31   0.001    -.0600625   -.0153559
     rcs():3 |     .01592   .0091416     1.74   0.082    -.0019973    .0338372
 time#M2[id] |          1          .        .       .            .           .
      M1[id] |          1          .        .       .            .           .
       _cons |   1.079191   .0798641    13.51   0.000     .9226599    1.235721
  sd(resid.) |   .3456885   .0067258                      .3327544    .3591254
-------------+----------------------------------------------------------------
id:          |            
      sd(M1) |   .9858939   .0421404                      .9066653    1.072046
      sd(M2) |   .1819288   .0127223                      .1586269    .2086537
 corr(M2,M1) |   .4340876   .0739756                       .278697    .5673296
------------------------------------------------------------------------------
\end{verbatim}

which shows evidence of positive correlation between intercept and
slope.

\subsection{User-defined model}\label{user-defined-model}

Now I'm going to show you how to replicate the previous model using your
own log likelihood function. You simply write your function in Mata,
which returns a \texttt{real\ matrix} containing the observation level
log likelihood contribution. This is the code for a Gaussian distributed
repsonse:

\begin{verbatim}
mata:
real matrix logl(M)
{
    y     = merlin_util_depvar(M)     //extract the response
    xb    = merlin_util_xzb(M)        //get the complex predictor
    sdre  = exp(merlin_util_ap(M,1))  //get the ancillary residual error std. dev.
    logl  = lnnormalden(y,xb,sdre)    //calculate ob. level log likelihood
    return(logl)                      //return result
}
end
\end{verbatim}

I get my dependent variable, my linear predictor, my one ancillary
parameter for the residual standard deviation (estimated on the log
scale), and use Mata's internal log normal density function. Now I
simply pass it to \texttt{merlin} as a \texttt{family(user)}, specifying
\texttt{nap(1)} for my residual standard deviation, and you can specify
anything you like in the complex predictor. Let's fit the previous
model,

\begin{verbatim}
. merlin (logb rcs(time, df(3) orthog) time#M2[id]@1 M1[id]@1,            ///
>                               family(user, llfunction(logl) nap(1))),   ///
>                               covariance(unstructured) nolog
variables created for model 1, component 1: _cmp_1_1_1 to _cmp_1_1_3

Fitting fixed effects model:

Fitting full model:

Mixed effects regression model                  Number of obs     =      1,945
Log likelihood =  -1518.484
------------------------------------------------------------------------------
             |      Coef.   Std. Err.      z    P>|z|     [95% Conf. Interval]
-------------+----------------------------------------------------------------
logb:        |            
     rcs():1 |   .5975687    .044098    13.55   0.000     .5111382    .6839992
     rcs():2 |  -.0377092    .011405    -3.31   0.001    -.0600625   -.0153559
     rcs():3 |     .01592   .0091416     1.74   0.082    -.0019973    .0338372
 time#M2[id] |          1          .        .       .            .           .
      M1[id] |          1          .        .       .            .           .
       _cons |   1.079191   .0798641    13.51   0.000     .9226599    1.235721
        ap:1 |  -1.062217   .0194562   -54.60   0.000    -1.100351   -1.024084
-------------+----------------------------------------------------------------
id:          |            
      sd(M1) |   .9858939   .0421404                      .9066653    1.072046
      sd(M2) |   .1819288   .0127223                      .1586269    .2086537
 corr(M2,M1) |   .4340876   .0739756                       .278697    .5673296
------------------------------------------------------------------------------
\end{verbatim}

which gives us identical results, as expected. The potential of this
sort of implementation is huge.

\subsection{Survival/time-to-event
analysis}\label{survivaltime-to-event-analysis}

Many standard time-to-event models are available in \texttt{merlin},
including most of those in \texttt{streg}, and some more flexible
distributions are also inbuilt. This includes the Royston-Parmar model,
and a log hazard scale equivalent model, both using restricted cubic
splines to model the baseline. To fit a survival model with
\texttt{merlin}, we simply add the \texttt{failure(varname)} option, to
specify the event indicator.

\subsubsection{Weibull proportional hazards
model}\label{weibull-proportional-hazards-model}

Let's start with a simple Weibull proportional hazards model, adjusting
for treatment:

\begin{verbatim}
. merlin (stime trt, family(weibull, failure(died)))

Fitting full model:

Iteration 0:   log likelihood = -2000.3067  
Iteration 1:   log likelihood = -512.23492  
Iteration 2:   log likelihood = -511.85192  
Iteration 3:   log likelihood = -511.84742  
Iteration 4:   log likelihood = -511.84742  

Mixed effects regression model                  Number of obs     =        312
Log likelihood = -511.84742
------------------------------------------------------------------------------
             |      Coef.   Std. Err.      z    P>|z|     [95% Conf. Interval]
-------------+----------------------------------------------------------------
stime:       |            
         trt |  -.0004536    .169053    -0.00   0.998    -.3317914    .3308841
       _cons |  -2.815926   .2054485   -13.71   0.000    -3.218597   -2.413254
  log(gamma) |   .0740757   .0752114     0.98   0.325    -.0733359    .2214873
------------------------------------------------------------------------------
\end{verbatim}

Simple, yet effective. Let's explore some non-standard models.

\subsubsection{Spline-based survival
model}\label{spline-based-survival-model}

The Royston-Parmar model is now widely used, using restricted cubic
splines to model the baseline log cumulative hazard function, and any
time-dependent effects:

\begin{verbatim}
. merlin (stime trt, family(rp, failure(died) df(3)))
variables created: _rcs1_1 to _rcs1_3

Fitting full model:

Iteration 0:   log likelihood = -488.26676  
Iteration 1:   log likelihood = -356.60857  
Iteration 2:   log likelihood = -354.86415  
Iteration 3:   log likelihood = -354.85331  
Iteration 4:   log likelihood = -354.85331  

Mixed effects regression model                  Number of obs     =        312
Log likelihood = -354.85331
------------------------------------------------------------------------------
             |      Coef.   Std. Err.      z    P>|z|     [95% Conf. Interval]
-------------+----------------------------------------------------------------
stime:       |            
         trt |   .0001164   .1690663     0.00   0.999    -.3312475    .3314803
       _cons |  -1.088126   .1258163    -8.65   0.000    -1.334721   -.8415307
------------------------------------------------------------------------------
\end{verbatim}

Note the spline coefficients aren't shown by default (they're
essentially uninterpretable). You can show all the parameters by using
\texttt{ml\ display}.

\subsubsection{Assess
proportional-hazards}\label{assess-proportional-hazards}

Let's assess proportional hazards in the effect of treatment by forming
an interaction between treatment and log time:

\begin{verbatim}
. merlin (stime trt trt#fp(stime, powers(0)), family(rp, failure(died) df(3)) ///
>                                             timevar(stime))
variables created: _rcs1_1 to _rcs1_3
variables created for model 1, component 2: _cmp_1_2_1 to _cmp_1_2_1

Fitting full model:

Iteration 0:   log likelihood = -488.26676  
Iteration 1:   log likelihood = -356.33217  (not concave)
Iteration 2:   log likelihood = -354.88597  (not concave)
Iteration 3:   log likelihood = -354.58729  
Iteration 4:   log likelihood = -354.48331  
Iteration 5:   log likelihood = -354.48326  
Iteration 6:   log likelihood = -354.48326  

Mixed effects regression model                  Number of obs     =        312
Log likelihood = -354.48326
------------------------------------------------------------------------------
             |      Coef.   Std. Err.      z    P>|z|     [95% Conf. Interval]
-------------+----------------------------------------------------------------
stime:       |            
         trt |  -.2856643   .3731528    -0.77   0.444     -1.01703    .4457017
    trt#fp() |   .1395046    .162367     0.86   0.390    -.1787289     .457738
       _cons |  -1.058688   .1293423    -8.19   0.000    -1.312194   -.8051815
------------------------------------------------------------------------------
\end{verbatim}

we had to specify our \texttt{timevar()} so \texttt{merlin} knows to
handle it differently. Equivalently, we could've used the \texttt{rcs()}
element type:

\begin{verbatim}
merlin (stime trt trt#rcs(stime, df(1) log),  family(rp, failure(died) df(3)) ///
                                              timevar(stime))
\end{verbatim}

\subsubsection{Add a non-linear effect}\label{add-a-non-linear-effect}

We can keep building this model, by investigating the effect of age,
modelled flexibly using fractional polynomials:

\begin{verbatim}
. merlin (stime trt                                   ///
>               fp(age, pow(1 1))                     ///
>               trt#fp(stime, powers(0))              ///
>               , family(rp, failure(died) df(3))     ///
>               timevar(stime))                       ///
>               , nolog
variables created: _rcs1_1 to _rcs1_3
variables created for model 1, component 2: _cmp_1_2_1 to _cmp_1_2_2
variables created for model 1, component 3: _cmp_1_3_1 to _cmp_1_3_1

Fitting full model:

Mixed effects regression model                  Number of obs     =        312
Log likelihood =  -339.6019
------------------------------------------------------------------------------
             |      Coef.   Std. Err.      z    P>|z|     [95% Conf. Interval]
-------------+----------------------------------------------------------------
stime:       |            
         trt |   -.435578   .3744746    -1.16   0.245    -1.169535    .2983787
      fp():1 |   .0168284   .2551338     0.07   0.947    -.4832246    .5168815
      fp():2 |    .005881   .0513732     0.11   0.909    -.0948086    .1065707
    trt#fp() |   .1340345   .1639652     0.82   0.414    -.1873314    .4554003
       _cons |   -3.04788   2.695243    -1.13   0.258     -8.33046    2.234699
------------------------------------------------------------------------------
\end{verbatim}

and for completeness, we can of course investigate whether proportional
hazards is valid for the age function,

\begin{verbatim}
. merlin (stime trt                                     ///
>               fp(age, pow(1 1))                       ///
>               fp(age, pow(1 1))#fp(stime, pow(0))     ///
>               trt#fp(stime, powers(0))                ///
>               , family(rp, failure(died) df(3))       ///
>               timevar(stime))                         ///
>               , nolog
variables created: _rcs1_1 to _rcs1_3
variables created for model 1, component 2: _cmp_1_2_1 to _cmp_1_2_2
variables created for model 1, component 3: _cmp_1_3_1 to _cmp_1_3_2
variables created for model 1, component 4: _cmp_1_4_1 to _cmp_1_4_1

Fitting full model:

Mixed effects regression model                  Number of obs     =        312
Log likelihood = -338.69894
------------------------------------------------------------------------------
             |      Coef.   Std. Err.      z    P>|z|     [95% Conf. Interval]
-------------+----------------------------------------------------------------
stime:       |            
         trt |  -.4744271   .3734131    -1.27   0.204    -1.206303    .2574492
      fp():1 |   .6637168   .3303349     2.01   0.045     .0162722    1.311161
      fp():2 |  -.1204047   .0662931    -1.82   0.069    -.2503369    .0095274
 fp()#fp():1 |  -.3097024   .0167436   -18.50   0.000    -.3425192   -.2768856
 fp()#fp():2 |   .0604016   .0036997    16.33   0.000     .0531504    .0676529
    trt#fp() |   .1554519   .1637138     0.95   0.342    -.1654214    .4763251
       _cons |  -4.792055   3.470469    -1.38   0.167    -11.59405     2.00994
------------------------------------------------------------------------------
\end{verbatim}

\subsubsection{Delayed
entry/left-truncation}\label{delayed-entryleft-truncation}

We can allow for delayed entry/left-truncation by using the
\texttt{ltruncated()} option. Since we have no delayed entry in this
dataset, I simulate one as an illustration:

\begin{verbatim}
. set seed 42590

. gen t0 = runiform()*stime*0.5
(1,633 missing values generated)

. merlin (stime trt                                               ///
>               fp(age, pow(1 1))                                 ///
>               fp(age, pow(1 1))#fp(stime, pow(0))               ///
>               trt#fp(stime, powers(0))                          ///
>               , family(rp, failure(died) df(3) ltruncated(t0))  ///
>               timevar(stime))                                   ///
>               , nolog
variables created: _rcs1_1 to _rcs1_3
variables created for model 1, component 2: _cmp_1_2_1 to _cmp_1_2_2
variables created for model 1, component 3: _cmp_1_3_1 to _cmp_1_3_2
variables created for model 1, component 4: _cmp_1_4_1 to _cmp_1_4_1

Fitting full model:

Mixed effects regression model                  Number of obs     =        312
Log likelihood = -286.65131
------------------------------------------------------------------------------
             |      Coef.   Std. Err.      z    P>|z|     [95% Conf. Interval]
-------------+----------------------------------------------------------------
stime:       |            
         trt |  -.5234664   .5618975    -0.93   0.352    -1.624765    .5778326
      fp():1 |   .3022418   .3293594     0.92   0.359    -.3432908    .9477743
      fp():2 |  -.0488895   .0668201    -0.73   0.464    -.1798545    .0820754
 fp()#fp():1 |  -.1271631   .0121461   -10.47   0.000     -.150969   -.1033571
 fp()#fp():2 |   .0244367    .003049     8.01   0.000     .0184607    .0304126
    trt#fp() |   .1226594   .1794447     0.68   0.494    -.2290458    .4743645
       _cons |   -3.21389   3.453738    -0.93   0.352    -9.983092    3.555312
------------------------------------------------------------------------------
\end{verbatim}

\subsection{Competing risks}\label{competing-risks}

Now I'll look at a multiple outcome survival model, for example the
competing risks setting. This is done by specifying cause-specific
hazard models. Remember our data structure is simplest in wide format,
and within a competing risks setting, our survival time is either
censoring or event time, regardless of the event, we simply need
cause-specific event indicators. Since this dataset only has all-cause
mortality, a random assign half of them to represent death from cancer,
\texttt{cancer}, and the other half to death from other causes,
\texttt{other}. I can then model each cause-specific hazard in any way
we like, for example:

\begin{verbatim}
. gen cancer = 1 if died==1 & runiform()<0.5
(1,870 missing values generated)

. gen other = 1 if died==1 & cancer!=1
(1,880 missing values generated)

. replace cancer = 0 if died==0 | other==1
(237 real changes made)

. replace other = 0 if died==0 | cancer==1
(247 real changes made)

. merlin  (stime trt , family(rcs, failure(cancer) df(3)))    ///
>         (stime trt , family(rp, failure(other) df(3)))      ///
>         , nolog
variables created: _rcs1_1 to _rcs1_3
variables created: _rcs2_1 to _rcs2_3

Fitting full model:

Mixed effects regression model                  Number of obs     =        312
Log likelihood = -449.59187
------------------------------------------------------------------------------
             |      Coef.   Std. Err.      z    P>|z|     [95% Conf. Interval]
-------------+----------------------------------------------------------------
stime:       |            
         trt |   .0517294   .2311444     0.22   0.823    -.4013053    .5047641
       _cons |  -3.248855   .1792481   -18.12   0.000    -3.600174   -2.897535
-------------+----------------------------------------------------------------
stime:       |            
         trt |  -.0618846   .2481256    -0.25   0.803    -.5482019    .4244326
       _cons |   -1.85768   .1837267   -10.11   0.000    -2.217778   -1.497582
------------------------------------------------------------------------------
\end{verbatim}

Let's calculate the cumulative incidence functions for each cause using
the \texttt{predict} tools. I'll generate a time variable, and calculate
the predictions for a patient in the treated group.

\begin{verbatim}
. range tvar 0 10 100
(1,845 missing values generated)

. predict cif1, cif outcome(1) causes(1 2) timevar(tvar) at(trt 1)
(1846 missing values generated)

. predict cif2, cif outcome(2) causes(1 2) timevar(tvar) at(trt 1)
(1846 missing values generated)
\end{verbatim}

By default, \texttt{causes()} will include all models, but I'm being
explicit here for clarity. To create a stacked plot, we need to add
together the two CIFs, and then use \texttt{area} graphs:

\begin{verbatim}
. gen totalcif1 = cif1 + cif2

. twoway (area totalcif1 tvar)(area cif2 tvar), name(g1,replace)        ///
>           xtitle("Time since entry") ytitle("Cumulative incidence")   ///
>           title("Treated group") legend(cols(1)                       ///
>           order(1 "Prob. of death due to cancer"                      ///
>           2 "Prob. of death due to other causes"))                    ///
>           ylabel(,angle(h) format(%2.1f)) ylabel(0(0.1)1)             ///
>           plotregion(margin(zero))
\end{verbatim}

\begin{figure}[!h]
  \centering
  \includegraphics[width=0.7\textwidth]{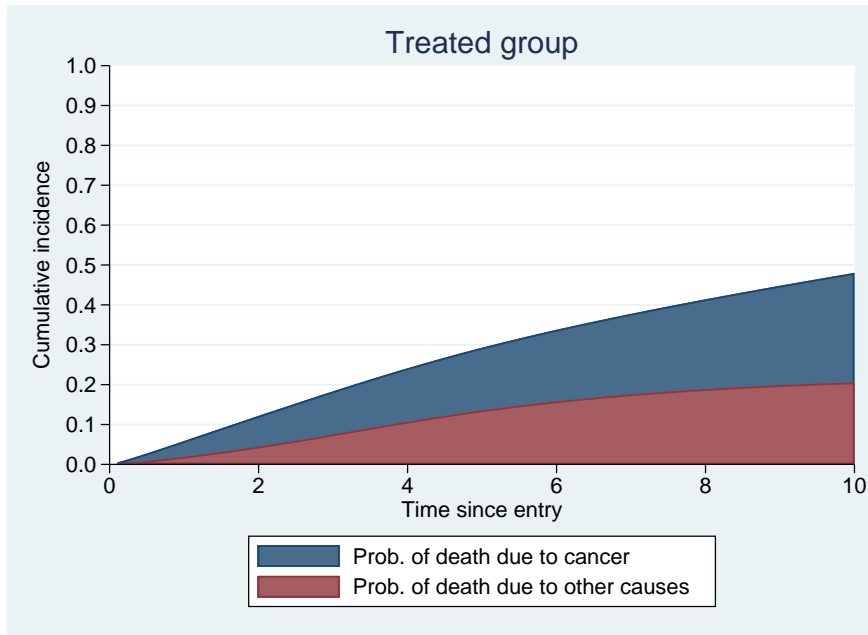}
  \caption{Stacked cumulative incidence functions}
\end{figure}

\subsection{A multiple-outcome/multivariate
model}\label{a-multiple-outcomemultivariate-model}

Now I'll move into the field of joint-longitudinal survival models.
Joint models were first proposed by linking the submodels through the
random effects only, such as:

\begin{verbatim}
. merlin  (stime trt M1[id]   , family(weibull, failure(died)))   ///
>         (logb time M1[id]@1 , family(gaussian))

Fitting fixed effects model:

Fitting full model:

Iteration 0:   log likelihood = -3095.8911  
Iteration 1:   log likelihood =  -2494.924  
Iteration 2:   log likelihood = -2367.2026  
Iteration 3:   log likelihood =  -2310.837  
Iteration 4:   log likelihood = -2310.5375  
Iteration 5:   log likelihood = -2310.5374  

Mixed effects regression model                  Number of obs     =      1,945
Log likelihood = -2310.5374
------------------------------------------------------------------------------
             |      Coef.   Std. Err.      z    P>|z|     [95% Conf. Interval]
-------------+----------------------------------------------------------------
stime:       |            
         trt |   .0231889   .1749004     0.13   0.895    -.3196096    .3659873
      M1[id] |   1.233915   .1033909    11.93   0.000     1.031273    1.436557
       _cons |    -3.8462   .2687252   -14.31   0.000    -4.372892   -3.319509
  log(gamma) |   .4221851   .0711164     5.94   0.000     .2827996    .5615706
-------------+----------------------------------------------------------------
logb:        |            
        time |   .0976615   .0043191    22.61   0.000     .0891962    .1061268
      M1[id] |          1          .        .       .            .           .
       _cons |    .577775   .0649576     8.89   0.000     .4504604    .7050897
  sd(resid.) |   .4912122   .0085843                      .4746722    .5083284
-------------+----------------------------------------------------------------
id:          |            
      sd(M1) |   1.107264   .0470596                      1.018766     1.20345
------------------------------------------------------------------------------
\end{verbatim}

By including my random intercept \texttt{M1} in the complex predictor
for the survival outcome, and \emph{not} specifying a constraint, it
will give me an \emph{association} parameter, for the relationship
between the subject-specific random intercept of the biomaker, and
survival. Alternatively, and now more commonly, we can use the
\texttt{EV{[}{]}} element type to link the time-dependent expected value
of the biomarker, directly to the risk of event. Given we are modelling
time, the \texttt{timevar()} must be specified in both submodels, as it
will be integrated over in the survival model likelihood contribution.

\begin{verbatim}
. merlin  (stime trt EV[logb]             , family(weibull, failure(died))    ///
>                                                           timevar(stime))   ///
>         (logb fp(time, pow(1)) M1[id]@1 , family(gaussian) timevar(time))   ///
>         , nolog
variables created for model 2, component 1: _cmp_2_1_1 to _cmp_2_1_1

Fitting fixed effects model:

Fitting full model:

Mixed effects regression model                  Number of obs     =      1,945
Log likelihood =  -2307.676
------------------------------------------------------------------------------
             |      Coef.   Std. Err.      z    P>|z|     [95% Conf. Interval]
-------------+----------------------------------------------------------------
stime:       |            
         trt |   .0218858   .1753999     0.12   0.901    -.3218916    .3656632
        EV[] |   1.279504   .1064773    12.02   0.000     1.070812    1.488196
       _cons |  -4.544086   .2929153   -15.51   0.000    -5.118189   -3.969982
  log(gamma) |   .1816336   .0725664     2.50   0.012      .039406    .3238612
-------------+----------------------------------------------------------------
logb:        |            
        fp() |   .0981289   .0043134    22.75   0.000     .0896748     .106583
      M1[id] |          1          .        .       .            .           .
       _cons |   .5775447   .0649697     8.89   0.000     .4502065     .704883
  sd(resid.) |   .4912633    .008587                      .4747181    .5083852
-------------+----------------------------------------------------------------
id:          |            
      sd(M1) |   1.107605   .0470632                      1.019099    1.203797
------------------------------------------------------------------------------
\end{verbatim}

There's no limit in how many and what form of association we specify,
for example, we can also link the cumulative value of the biomarker
(this of it as cumulative exposure), directly to survival as well.

\begin{verbatim}
. merlin  (stime trt EV[logb] iEV[logb]   , family(weibull, failure(died))    ///
>                                                           timevar(stime))   ///
>         (logb fp(time, pow(1)) M1[id]@1 , family(gaussian) timevar(time))   ///
>         , nolog
variables created for model 2, component 1: _cmp_2_1_1 to _cmp_2_1_1

Fitting fixed effects model:

Fitting full model:

Mixed effects regression model                  Number of obs     =      1,945
Log likelihood = -2306.8166
------------------------------------------------------------------------------
             |      Coef.   Std. Err.      z    P>|z|     [95% Conf. Interval]
-------------+----------------------------------------------------------------
stime:       |            
         trt |   .0504001   .1769845     0.28   0.776    -.2964832    .3972833
        EV[] |   1.454447   .1759923     8.26   0.000     1.109508    1.799385
       iEV[] |   -.039734   .0309956    -1.28   0.200    -.1004842    .0210162
       _cons |  -4.897317   .4150173   -11.80   0.000    -5.710736   -4.083898
  log(gamma) |   .2881265   .1054815     2.73   0.006     .0813864    .4948665
-------------+----------------------------------------------------------------
logb:        |            
        fp() |   .0982338   .0043145    22.77   0.000     .0897775      .10669
      M1[id] |          1          .        .       .            .           .
       _cons |    .578134   .0650171     8.89   0.000     .4507029    .7055651
  sd(resid.) |   .4912445   .0085862                      .4747009    .5083647
-------------+----------------------------------------------------------------
id:          |            
      sd(M1) |   1.108471   .0471066                      1.019885    1.204753
------------------------------------------------------------------------------
\end{verbatim}

Non-linearities in the association, and time-dependent effects can also
be specified,

\begin{verbatim}
. merlin  (stime trt EV[logb] EV[logb]#fp(stime, pow(0))                            ///
>                                 , family(weibull, failure(died)) timevar(stime))  ///
>         (logb fp(time, pow(1)) M1[id]@1 , family(gaussian) timevar(time))         ///
>         , nolog
variables created for model 1, component 3: _cmp_1_3_1 to _cmp_1_3_1
variables created for model 2, component 1: _cmp_2_1_1 to _cmp_2_1_1

Fitting fixed effects model:

Fitting full model:

Mixed effects regression model                  Number of obs     =      1,945
Log likelihood = -2306.9583
------------------------------------------------------------------------------
             |      Coef.   Std. Err.      z    P>|z|     [95% Conf. Interval]
-------------+----------------------------------------------------------------
stime:       |            
         trt |   .0474124    .176971     0.27   0.789    -.2994445    .3942693
        EV[] |   1.426935   .1685864     8.46   0.000     1.096512    1.757359
   EV[]#fp() |  -.1238257    .103456    -1.20   0.231    -.3265958    .0789443
       _cons |  -5.039479   .5227427    -9.64   0.000    -6.064036   -4.014922
  log(gamma) |   .3605683   .1549538     2.33   0.020     .0568644    .6642722
-------------+----------------------------------------------------------------
logb:        |            
        fp() |   .0980068   .0043171    22.70   0.000     .0895455    .1064682
      M1[id] |          1          .        .       .            .           .
       _cons |   .5792938   .0650543     8.90   0.000     .4517897    .7067979
  sd(resid.) |   .4912286   .0085855                      .4746863    .5083473
-------------+----------------------------------------------------------------
id:          |            
      sd(M1) |   1.108874   .0471116                      1.020277    1.205165
------------------------------------------------------------------------------
\end{verbatim}

\subsection{A final model}\label{a-final-model}

So now I'm going to bring together a lot of the previous models into one
meta-\texttt{merlin}-model, if you will. This is purely for illustrative
purposes, but hopefully gives you an idea of just how flexible
\texttt{merlin} can be. I'm now going to generate a binary repeatedly
measured variable, \texttt{catpro}, which is \texttt{prothrombin} merely
categorised into above and below 12 (this is clearly an unnecessary
thing to do, but purely for illustrative purposes). I will also use the
artificial competing risks outcomes created above, and so bring together
a joint model for two repeatedly measured outcomes, one continuous and
one binary, with cause-specific competing risks survival models. Each
cause-specific hazard model will have a different distribution, and we
can allow for time-dependent effects/non-proportional hazards.

\begin{verbatim}
. gen byte catpro = prothrombin > 12
\end{verbatim}

\begin{verbatim}
. merlin  (stime  trt                                     ///
>                 M2[id] M1[id]                           ///
>                 , family(rcs, failure(cancer) df(3))    ///
>                 timevar(stime))                         ///
>         (stime  trt trt#fp(stime, pow(0))               ///
>                 EV[logb] M1[id]                         ///
>                 , family(weibull, failure(other))       ///
>                 timevar(stime))                         ///
>         (logb   rcs(time, df(3) orthog)                 ///
>                 M1[id]@1                                ///
>                 , family(gaussian)                      ///
>                 timevar(time))                          ///
>         (catpro fp(time, powers(1))                     ///
>                 M2[id]@1                                ///
>                 , family(bernoulli)                     ///
>                 timevar(time))                          ///
>         , covariance(unstructured) nolog
variables created: _rcs1_1 to _rcs1_3
variables created for model 2, component 2: _cmp_2_2_1 to _cmp_2_2_1
variables created for model 3, component 1: _cmp_3_1_1 to _cmp_3_1_3
variables created for model 4, component 1: _cmp_4_1_1 to _cmp_4_1_1

Fitting fixed effects model:

Fitting full model:

Mixed effects regression model                  Number of obs     =      1,945
Log likelihood = -2827.9382
------------------------------------------------------------------------------
             |      Coef.   Std. Err.      z    P>|z|     [95% Conf. Interval]
-------------+----------------------------------------------------------------
stime:       |            
         trt |   .0843121   .2585119     0.33   0.744    -.4223619    .5909861
      M2[id] |   .5877474   .1423448     4.13   0.000     .3087568     .866738
      M1[id] |   .6124566    .231844     2.64   0.008     .1580507    1.066863
       _cons |  -3.291482   .2618039   -12.57   0.000    -3.804608   -2.778356
-------------+----------------------------------------------------------------
stime:       |            
         trt |  -.0431231   .4161423    -0.10   0.917    -.8587469    .7725008
    trt#fp() |  -.0008662   .2759082    -0.00   0.997    -.5416363    .5399039
        EV[] |   1.079582   1.306597     0.83   0.409    -1.481301    3.640466
      M1[id] |   .1642467   1.280222     0.13   0.898    -2.344943    2.673436
       _cons |  -5.311877   .8722111    -6.09   0.000     -7.02138   -3.602375
  log(gamma) |   .2798372   .2963124     0.94   0.345    -.3009244    .8605988
-------------+----------------------------------------------------------------
logb:        |            
     rcs():1 |   .3046816   .0132337    23.02   0.000      .278744    .3306192
     rcs():2 |   .0678822   .0118585     5.72   0.000       .04464    .0911244
     rcs():3 |   .0136599   .0114777     1.19   0.234     -.008836    .0361558
      M1[id] |          1          .        .       .            .           .
       _cons |   .8969961   .0647674    13.85   0.000     .7700542    1.023938
  sd(resid.) |    .486531   .0085215                      .4701126    .5035228
-------------+----------------------------------------------------------------
catpro:      |            
        fp() |   .5194474   .0453455    11.46   0.000     .4305718    .6083229
      M2[id] |          1          .        .       .            .           .
       _cons |  -4.618131   .3378236   -13.67   0.000    -5.280253   -3.956009
-------------+----------------------------------------------------------------
id:          |            
      sd(M1) |   1.115666   .0473025                      1.026702    1.212338
      sd(M2) |   2.878879   .2668173                      2.400674     3.45234
 corr(M2,M1) |   .7836109    .034906                      .7051134    .8431351
------------------------------------------------------------------------------
\end{verbatim}

There's four outcome models, and two random effects, so it takes some
time, but we get there in about 16 minutes on my laptop. It's trivial to
extend this model, for example adding random slopes, or different ways
of linking the submodels, or incorporating non-linear effects, be they
for continuous covariates or time.

\section{Discussion}\label{discussion}

\label{sec:disc}

I have given a broad overview of \texttt{merlin}'s capabilities, and
potential, in the field of data analysis. I've described the fundamental
syntax, and through worked examples, illustrated some of the areas of
statistical modelling that can be applied using \texttt{merlin}. I'll
finish with some final thoughts.

\subsection{The curse of generality}\label{the-curse-of-generality}

When implementing a software package that can do many things, one of the
challenging tasks is to balance said generality, with computational
speed. \texttt{merlin} is not the quickest. It's current implementation
is a \texttt{gf0} evaluator, which means it uses Stata's internal finite
difference routines within the \texttt{ml} engine to calculate the score
and Hessian, which means a lot of calls to the evaluator program.
Furthermore, \texttt{merlin} has to cover a lot of different settings
and options, which inevitably mean a lot of storing information within
its main object. Now a lot of that is alleviated through use of
\texttt{pointers} and other clever things, but it will still result in
some overheads. You will get speed gains if you implement a specific
command, from scratch, designed for a specific setting.

\subsection{Shell commands}\label{shell-commands}

Firstly, the syntax of \texttt{merlin} is not the simplest. This is of
course because it has to accommodate a lot of different options and
techniques. This opens up more room to go wrong. Secondly, it is rather
challenging to obtain good starting values for a command that can fit
anything, and good starting values can be crucial to improve
convergence.

This motivates the writing of shell commands, i.e.~ado files
specifically written to handle specific classes of \texttt{merlin}
models. This allows a much simpler and cleaner syntax, and the ability
to hard code initial value fitting routines. Under the hood, and
unbeknownst to most users, the shell command will call \texttt{merlin}.
I'm working on a few of these.

\subsection{Concluding remarks}\label{concluding-remarks}

There is a multitude of future directions to take \texttt{merlin} in.
Some of which include; adding analytic derivatives for computational
speed gains, extending the allowed random effects distributions to allow
things like mixtures of Gaussians, and providing more tools for
postestimation, including dynamic prediction capabilities.

The latest stable version of \texttt{merlin} can be installed by typing
\texttt{ssc\ install\ merlin} in Stata. The development version can be
installed using:

\begin{verbatim}
net install merlin, from(https://www.mjcrowther.co.uk/code/merlin)
\end{verbatim}

I hope you find \texttt{merlin} useful.

\section*{About the author}

Michael J. Crowther is a Lecturer in Biostatistics at the University of
Leicester. He works heavily in methods and software development,
particularly in the field of survival analysis. He is currently part
funded by a MRC New Investigator Research Grant (MR/P015433/1).

\renewcommand\refname{References}
\bibliography{merlin_tutorial_arxiv.bbl}

\end{document}